\documentclass[aps,prb,reprint,superscriptaddress]{revtex4-2}
\usepackage[T1]{fontenc}
\usepackage[utf8]{inputenc}
\usepackage{amsmath,amssymb}
\usepackage{graphicx}
\usepackage{xcolor}
\usepackage{microtype}

\usepackage{hyperref}

\hypersetup{
    colorlinks=true,
    linkcolor=blue,
    citecolor=blue,
    urlcolor=blue
}

\begin{document}

\title{Colossal magnetostriction effect in rare-earth orthoferrites}

\author{Moumita Das}
\affiliation{Saha Institute of Nuclear Physics, Homi Bhabha National Institute, 1/AF Bidhannagar, Kolkata 700064, India}
\author{Arup Ghosh}
\affiliation{Saha Institute of Nuclear Physics, Homi Bhabha National Institute, 1/AF Bidhannagar, Kolkata 700064, India}
\author{Moumita Nandi}
\affiliation{Department of Condensed Matter Physics and Materials Science, Tata Institute of Fundamental Research, Dr. Homi Bhabha Road, Colaba, Mumbai 400005, India}
\author{Arumugam Thamizhavel}
\affiliation{Department of Condensed Matter Physics and Materials Science, Tata Institute of Fundamental Research, Dr. Homi Bhabha Road, Colaba, Mumbai 400005, India}
\author{Dipten Bhattacharya}
\affiliation{Multiscale Microstructure and Mechanics of Materials Division, CSIR-Central Glass and Ceramic Research Institute, Kolkata 700032, India}
\author{Prabhat Mandal}
\email{prabhat.mandal@saha.ac.in}
\affiliation{Saha Institute of Nuclear Physics, Homi Bhabha National Institute, 1/AF Bidhannagar, Kolkata 700064, India}

\begin{abstract}
Within the entire gamut of magnetostrictive, piezomagnetic, and ferromagnetic shape memory alloy systems, the striction effect is found to vary from a few tens of parts per million to a few percent. Here, we report the observation of an unprecedentedly large magnetic-field-induced lattice strain along the $c$ axis (more than 60--100\% at 90 kOe field) in single crystals of orthorhombic $R$FeO$_3$ ($R$ = Dy, Ho) below their spin-reorientation transition temperature $T_{SR}$. However, the striction effect is two orders of magnitude smaller along the $a$ and $b$ axes. Like magnetostriction, the magnetodielectric effect is highly anisotropic and very large along the $c$ axis. Such gigantic striction and dielectric effects along the $c$ axis arise as a result of a magnetic-field-induced first-order structural phase transition which possibly stems from large spin-orbit coupling (and, thereby, enormous magnetocrystalline anisotropy). The observed colossal striction in rare-earth orthoferrites could open new opportunities for applications requiring large, reversible magnetic-field-induced strain.
\end{abstract}

\maketitle

\section{Introduction}
Because of magnetocrystalline anisotropy and different unit cell lengths along the crystallographic directions, a change in the domain structure and domain rotation under a magnetic field ($H$) leads to a change in the size of the magnetic sample as the magnetic moments reorient towards the applied-field direction to minimize the free energy. In these cases, magnetostriction \cite{Joule} varies quadratically with magnetization ($M$).  Large magnetostriction can also originate from reorientation, as against rotation of domains, and adaptability of the structure under applied magnetic field \cite{Chopra}. The piezomagnetic systems, on the other hand, exhibit linear coupling between mechanical strain and  magnetic field \cite{Romanov}. Although the strain remains within a few tens to hundreds of parts per million (ppm) in all these cases, magnetic shape memory alloys \cite{Kohl,Chmielus} exhibit a very large mechanical strain under a magnetic field. In these alloys, magnetic anisotropy energy is very large as compared to elastic energy, as a result, an enormous internal stress is generated which causes structural transformation. Taken together, across the entire range of magnetic materials, the striction effect turns out to vary from a few tens of ppm to 12\%.

Orthorhombic $R$FeO$_3$ ($R$=Dy, Ho) belongs to the rare-earth orthoferrite family, has attracted considerable attention in recent years due to their non-trivial spin texture \cite{Arty,Koba} and unusual physical properties such as  multiferroicity \cite{Toku,Naka,Tokuna},
gigantic magneto-optical phenomena \cite{Solo,Kahn}, ultrafast magnetic control\cite{Kim,Jong,Afan,Kuri,Vov}, topological magnons \cite{Karaki}, giant magnetocaloric effect \cite{Das}, etc. HoFeO$_3$ undergoes several magnetic transitions across 2--700 K from paramagnetic to antiferromagnetic $\Gamma_4$ at $\sim$650 K and spin reorientation ones such as $\Gamma_4$ $\rightarrow$ $\Gamma_1$ at $T_{SR}$=55 K, $\Gamma_1$ $\rightarrow$ $\Gamma_2$ at 35 K. DyFeO$_3$ also shows similar multiple transitions with spin reorientation transition near 50 K (Morin transition). In both HoFeO$_3$ and DyFeO$_3$, rare-earth moments order at 4 K. However, unlike HoFeO$_3$, the spin reorientation of Fe in DyFeO$_3$ occurs abruptly due to the stronger magnetocrystalline anisotropy of Dy than Ho. Though, $R$FeO$_3$ exhibit several interesting phenomena with applied magnetic field, structural properties under magnetic field have not been examined so far. Whether different phases could emerge under varying magnetic fields applied across the spin-reorientation transition needs to be examined in greater detail.  Remarkably, we discover that both HoFeO$_3$ and DyFeO$_3$ exhibit magnetic-field-induced first-order structural transition and unprecedentedly large mechanical strain and magnetodielectric effect along $c$ axis (of the order $\sim$60-100\% at 90 kOe along $c$ axis) over a wide range of temperature ($T$) and field. Both the field-induced transition and the strain are highly anisotropic.

\begin{figure*}[t]
\centering
\includegraphics[width=0.9\textwidth, trim=0cm 2cm 0cm 1cm, clip]{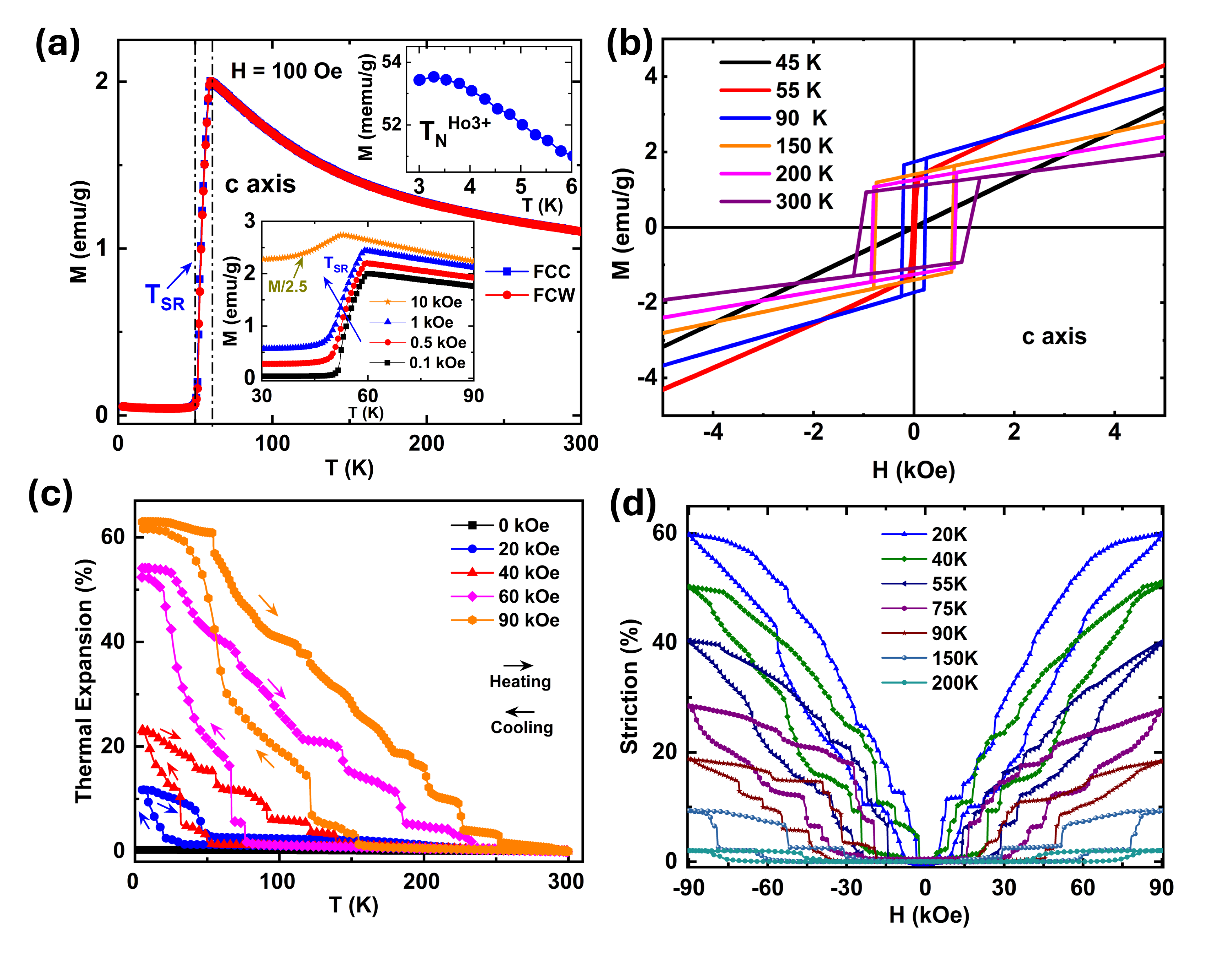}
\caption{(a) The main panel shows the magnetization ($M$) versus temperature ($T$) plot recorded by applying the field along c-axis and measuring the c-axis magnetization of HoFeO$_3$ crystal; the data obtained during the field-cooling-and-warming cycle of measurement (where $\sim$100 Oe was applied at room temperature while ramping down the temperature to $\sim$2.0 K followed by warming back to room temperature under field) are shown; the data clearly show the spin-reorientation transition of Fe at $T_{SR}$ $\approx$ 55 K; lower inset shows how the $T_{SR}$ varies with field across the range 0.1-10 kOe; $T_{SR}$ decreases rapidly with increasing field; the upper inset shows the antiferromagnetic ordering of Ho$^{3+}$ spins near 4 K in the magnetization data recorded along c-axis; (b) c-axis magnetization ($M$) versus field ($H$) hysteresis loops recorded within the low-field regime ($\pm$5 kOe) across the $T_{SR}$ are shown; interestingly, finite hysteresis and, therefore, coercivity could be observed above $T_{SR}$; the hysteresis collapses below $T_{SR}$ indicating precipitous drop in weak ferromagnetism; (c) the percentage change of thermal expansion ($(\Delta l/l)_c$) along c axis versus temperature at different magnetic fields applied along the c-axis of the crystal during cooling and heating; the $(\Delta l/l)_c$ rises dramatically below the corresponding $T_{SR}$ with large thermal hysteresis and as the temperature is decreased below $T_{SR}$ the step-like rise is observed, eventually resulting in gigantic thermal expansion at a temperature far below $T_{SR}$; $(\Delta l/l)_c$ reaches an unprecedented magnitude $>$60\% near 2 K under 90 kOe field; (d) the c-axis striction ($s$) versus magnetic field ($H$) loops - extracted from the thermal expansion under different magnetic field - at different temperatures around $T_{SR}$ are shown; a cascade of metamagnetic transitions with butterfly-like features could be observed; the $s-H$ data exhibit significant hysteresis as well.}
\label{fig:HFO_MST}
\end{figure*}

\begin{figure*}[t]
\centering
\includegraphics[width=0.9\textwidth]{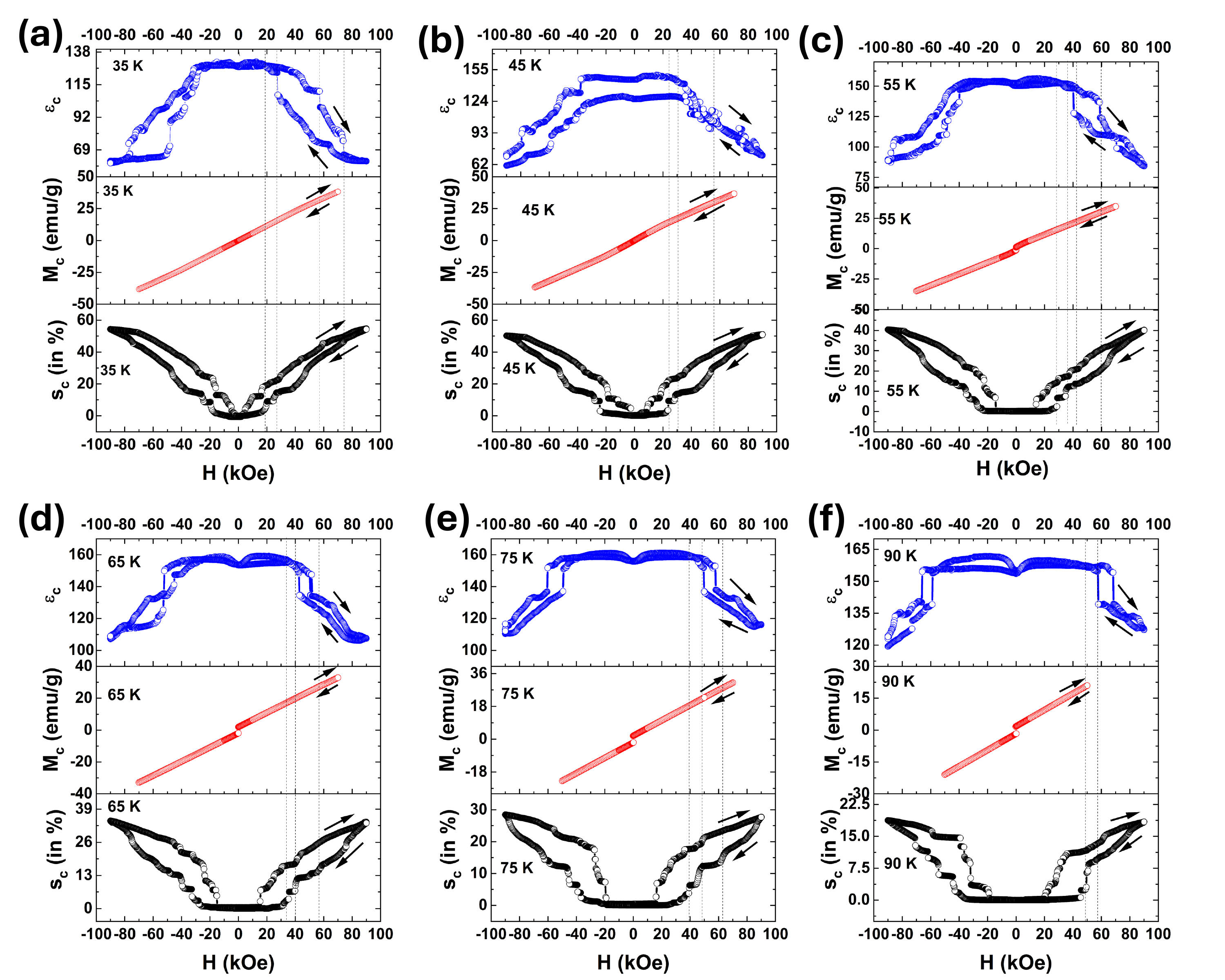}
\caption{The c-axis striction ($s$), magnetization ($M$), and dielectric constant ($\epsilon$) versus magnetic field patterns at different temperatures for HoFeO$_3$ are shown in this series of stack plots; importantly, the dielectric constant ($\epsilon$) exhibits significant magnetic field dependence at the selected temperatures and also step-like anomalous jumps at corresponding characteristic fields $H_C$ where the striction ($s$) exhibits similar features; $\epsilon$ drops by more than 65\% at 90 kOe field and 35 K; these results reveal a pronounced magnetodielectric response that is strongly correlated with the field-induced magnetostructural transitions; however, characteristic features in the magnetization versus field loops are absent possibly because of thermal broadening at higher temperatures; characteristic transition features in the $M-H$ hysteresis loops, however, could be observed at low temperature.}
\label{fig:HFO_MST_dielec}
\end{figure*}

\section{Results}
\subsection{Thermal expansion, magnetostriction and magnetodielectric response of HoFeO$_3$}
Figure \ref{fig:HFO_MST} summarizes the results of the measurement of $M$ versus $T$, $M$ versus  $H$, thermal expansion ($\Delta l/l$) and longitudinal striction ($s$) effect along $c$ axis, over a wide range of $H$ (applied parallel to the $c$-axis) and $T$ for HoFeO$_3$. $M$($T$) curve in Figure \ref{fig:HFO_MST}(a) shows that the spin-reorientation transition takes place at $T_{SR}$ $\sim$55 K and it shifts rapidly towards lower temperatures with the increase of magnetic field strength. The $M$($H$) curve of HoFeO$_3$ along the $c$ axis is typical of a canted antiferromagnet above $T_{SR}$ caused by the antisymmetric Dzyaloshinskii-Moriya (DM) exchange interaction between the antiferromagnetically ordered Fe$^{3+}$ spins (Figure \ref{fig:HFO_MST}(b)). Figure \ref{fig:HFO_MST}(c) depicts $T$ dependence of $\Delta l/l$ at different $H$. $\Delta l/l$ exhibits a very sharp two-step transition with large thermal hysteresis. This behavior clearly suggests that  HoFeO$_3$ undergoes a first-order structural phase transition under magnetic field. The value of $\Delta l/l$ increases rapidly with field and reaches more than 60\% at 90 kOe and 2 K. $H$ dependence of $s$ has been shown in Figure \ref{fig:HFO_MST}(d) for a few selected temperatures. $s$ increases rapidly with increasing $H$ and eventually reaches a gigantic magnitude - nearly 60\% at 20 K - under a magnetic field of $\sim$90 kOe. $s$($H$) exhibits characteristic features of the metamagnetic transitions with strong hysteresis between the increasing and decreasing fields. However, these features are very weak or nearly smeared in the $M-H$ loops.

\begin{figure*}[t]
\centering
\includegraphics[width=0.9\textwidth, trim=0cm 2cm 0cm 1cm, clip]{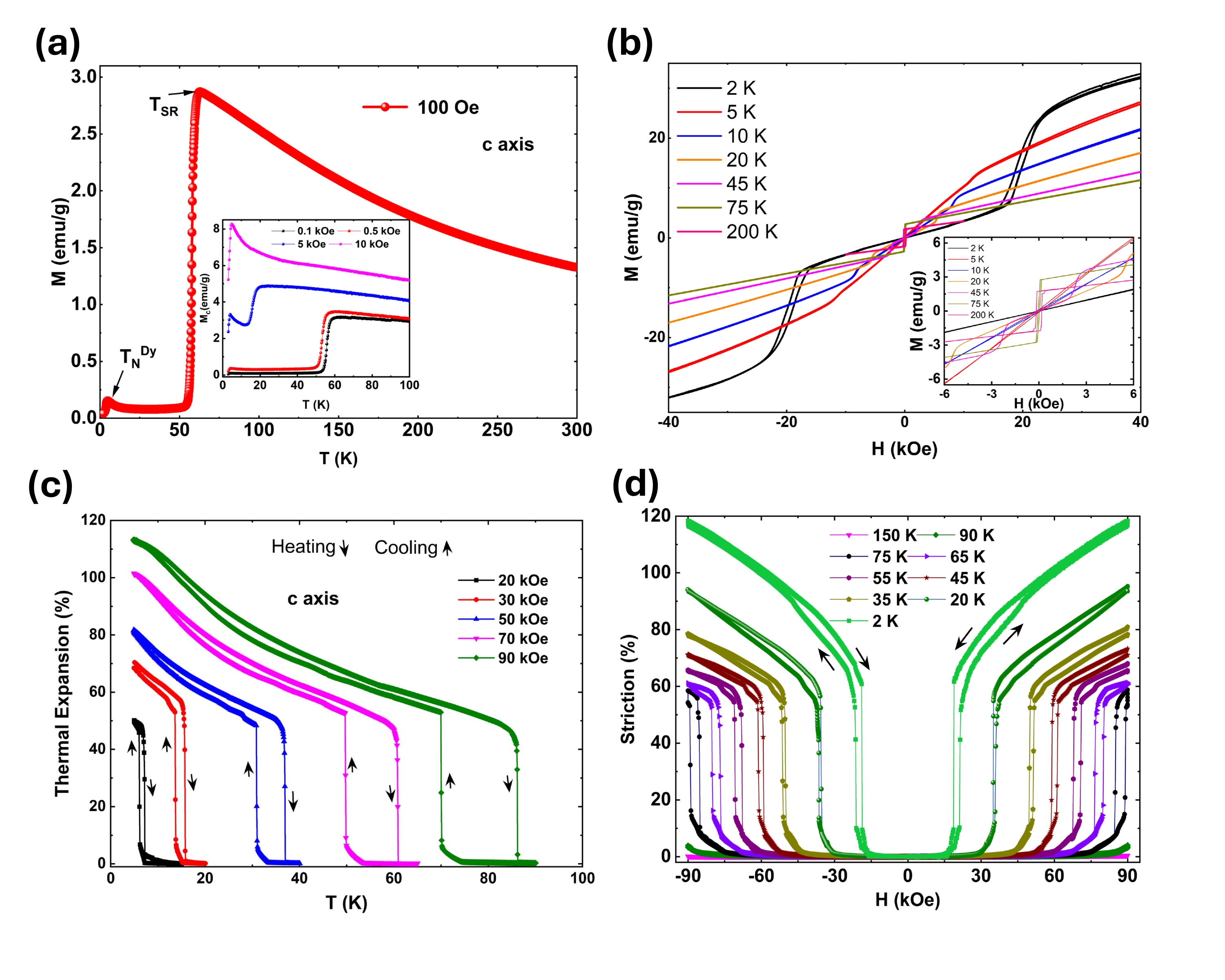}
\caption{(a) The main panel shows the c-axis magnetization ($M$) versus temperature ($T$) plot under 100 Oe magnetic field for DyFeO$_3$; the data were recorded during field-cooling-and-warming cycle where following field cooling from room temperature down to $\sim$2 K the temperature of the sample was ramped back under 100 Oe field to room temperature; the magnetization was recorded during the heating; the Fe spin-reorientation transition could be observed at $T_{Sr}$ $\approx$ 50 K; at even lower temperature near 4.2 K, antiferromagnetic ordering of Dy $^{3+}$ spins set in ($T_N^{Dy}$); inset shows the $M-T$ plots under different magnetic field 0.1-10 kOe; (b) the main panel shows the magnetic hysteresis loops at high fields across the corresponding $T_{SR}$; inset shows the blown-up portions of the hysteresis loops at lower field; like in HoFeO$_3$, in this case too, the hysteresis collapses at lower temperature below the respective $T_{SR}$;  (c) shows the thermal expansion ($(\Delta l/l)_c$) versus temperature ($T$) under different magnetic field across 20-90 kOe; cascade of metamagnetic jumps could be observed under different magnetic field; eventually at a lower temperature ($\sim$5 K) the $(\Delta l/l)_c$ reaches $\sim$115\%; (d) the c-axis striction ($s$) versus magnetic field ($H$) loops at different temperatures are shown here; in this case the magnetostriction reaches $\sim$120\%; this remarkably large magnetic field driven striction effect is, to our knowledge, among the largest field-induced strains reported so far; like in HoFeO$_3$, in this case too, hysteresis between the forward and reverse branches of the $s-H$ loop could be observed.}
\label{fig:DFO_MST}
\end{figure*}

Figure \ref{fig:HFO_MST_dielec} shows $H$ dependence of dielectric constant ($\epsilon$) along the $c$ axis at and below 90 K. Similar to $s$, $\epsilon$ exhibits a huge change with field and also the characteristic features of the metamagnetic transitions and hysteresis.  Away from the transition, $s$ appears to follow $M^2$ behavior.  Notably, the magnitude of $s$ and the nature of the $s-H$ loop of $c$ axis are different from that along $a$ and $b$ axes and so are the $H$ dependence of $M$ and $\epsilon$ (see Supplementary Information). The value of magnetostriction along these axes turns out to be quite small (varying within 0.1 to $\sim$3\%).  Also, the sharp features observed along $c$ axis are quite weak and smeared in the case of $a$ and $b$ axes.  $\epsilon$ also exhibits similar behavior along these axes.  In these directions too, the strain follows approximately $M^2$ dependence in the region far from the transition points.

\subsection{Thermal expansion and magnetostriction of DyFeO$_3$}

\begin{figure}[ht!]
\centering
\includegraphics[width=\columnwidth, trim=0cm 2cm 0cm 1cm, clip]{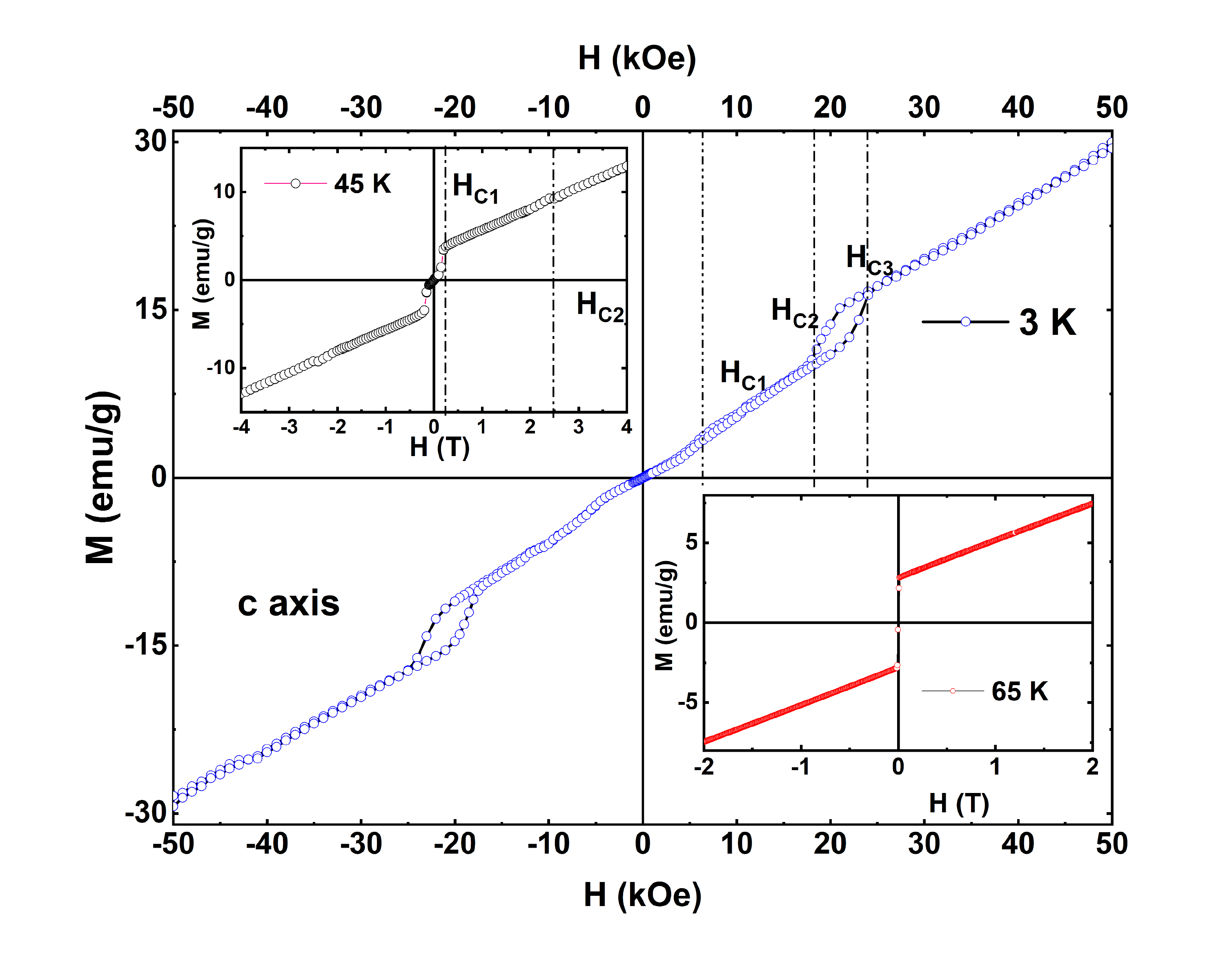}
\caption{\textbf{Low-temperature metamagnetic transitions in DyFeO$_3$.}
The main panel shows the $c$-axis magnetic hysteresis loop at $\sim$3 K.
Several anomalies corresponding to characteristic fields, $H_C$, are resolved in the $M-H$ loop. A clear hysteretic anomaly is observed at one of the fields, whereas no resolvable hysteresis
is observed at the remaining anomalies within the experimental
resolution. The upper and lower insets show representative $M-H$
hysteresis loops above and below $T_{SR}$, respectively.}
\label{fig:DFO_MH}
\end{figure}

Almost similar features have been observed in the case of DyFeO$_3$. Figure \ref{fig:DFO_MST}  summarizes the temperature and field dependences of magnetization, thermal expansion and magnetostriction along $c$ axis for DyFeO$_3$. In this case too, the spin-reorientation transition turns out to be $\sim$50 K and $T_{SR}$ shifts rapidly towards lower temperatures with field (Figure \ref{fig:DFO_MST}(a)). However, the transition is extremely sharp due to its first-order nature. Similar to HoFeO$_3$, $\Delta l/l$ for DyFeO$_3$ also exhibits strong thermal hysteresis (Figure \ref{fig:DFO_MST}(c)). The cascade of metamagnetic transitions yields even larger magnetostriction - more than 100\% - along $c$ axis below $T_{SR}$ (Figure \ref{fig:DFO_MST}(d)). In contrast, the striction effect turns out to be much smaller ($\sim$10\%) along $a$ and $b$ axes (see Supplementary Information). Therefore, the magnetostriction is highly anisotropic for both DyFeO$_3$ and HoFeO$_3$ compounds. Apart from the difference in the magnitude of the $c$-axis striction, HoFeO$_3$ exhibits multiple step-like transitions, whereas DyFeO$_3$ shows a single, extremely sharp transition. It is also important to point out that in DyFeO$_3$, the magnetic hysteresis loops measured at low temperatures (2--5 K) exhibit clear signatures of metamagnetic transitions, manifested as anomalies at characteristic fields (Figure \ref{fig:DFO_MH}). A clear hysteretic anomaly is resolved at one of the characteristic fields, whereas the other anomalies show no resolvable hysteresis within the experimental resolution. Similar but much weaker signatures are observed in the magnetization measured along the $a$ and $b$ axes. At higher temperatures, these features become progressively broadened by thermal effects.

\section{Discussion}

To examine the role of spin–orbit coupling in the magnetostriction, we have also measured the magnetic field dependence of striction in GdFeO$_3$ (see Supplementary Information figure 16). The magnetostriction along $c$ axis turns out to be about two orders of magnitude smaller as compared to DyFeO$_3$. This value of $s$ is comparable to that observed in HoFeO$_3$ and DyFeO$_3$ along $a$ and $b$ axes. Apart from the magnitude of magnetostriction, GdFeO$_3$ does not exhibit any magnetic-field-induced structural transition or metamagnetic transition. $M$($T$) and $M$($H$) curves do not show any feature and its nature is typical of a canted antiferromagnetic and $M$ is almost isotropic \cite{Das}. Unlike Ho and Dy, Gd$^{3+}$ has a spin-only moment; as a result, spin-orbit coupling in GdFeO$_3$ is very weak and magnetic ordering of Gd spin can be described by isotropic Heisenberg superexchange interactions.  On the other hand, due to strong spin-orbit coupling, Dy-moments with Ising-like nature align antiferromagnetically below $T_N^{Dy}$ and the Ising axis lies within the $ab$-plane \cite{Toku}. Another important difference is the absence of a spin-reorientation transition in GdFeO$_3$. Thus, our results confirm that there is a systematic pattern of variation of the magnitude of magnetostriction - from Dy to Ho to Gd - which correlates with the increasing importance of rare-earth orbital contributions and single-ion anisotropy from Gd to Ho and Dy.

Indeed, several recent reports have shown that electromagnetic properties of $R$FeO$_3$ along $c$ axis are  extremely sensitive to the external stimuli for compounds exhibiting strong spin-orbit coupling and spin-reorientation transition. The terahertz magnetic near field combined with femtosecond laser excitation has been able to induce over 80$\%$ of total magnetization along $c$ direction in ErFeO$_3$ by breaking the spin reorientation symmetry  \cite{Kuri}. Magnetic-field-induced multiferroicity and piezomagnetoelectric effect was observed in DyFeO$_3$ under uniaxial stress  \cite{Toku,Naka}. Large electric polarization along $c$ axis appears above 30 kOe by completely suppressing the spin-reorientation transition  \cite{Naka}. It has been suggested that the  interplay between two different magnetic ions, Fe$^{3+}$ and $R^{3+}$, leads to these remarkable functionalities. For example, the origin of multiferroicity and large electric polarization were explained by adapting the spin-dependent magnetostriction model to the nearest neighbor exchange coupling between the Fe and Dy moments.\cite{Toku}

\begin{figure}[ht!]
\centering
\includegraphics[width=\columnwidth]{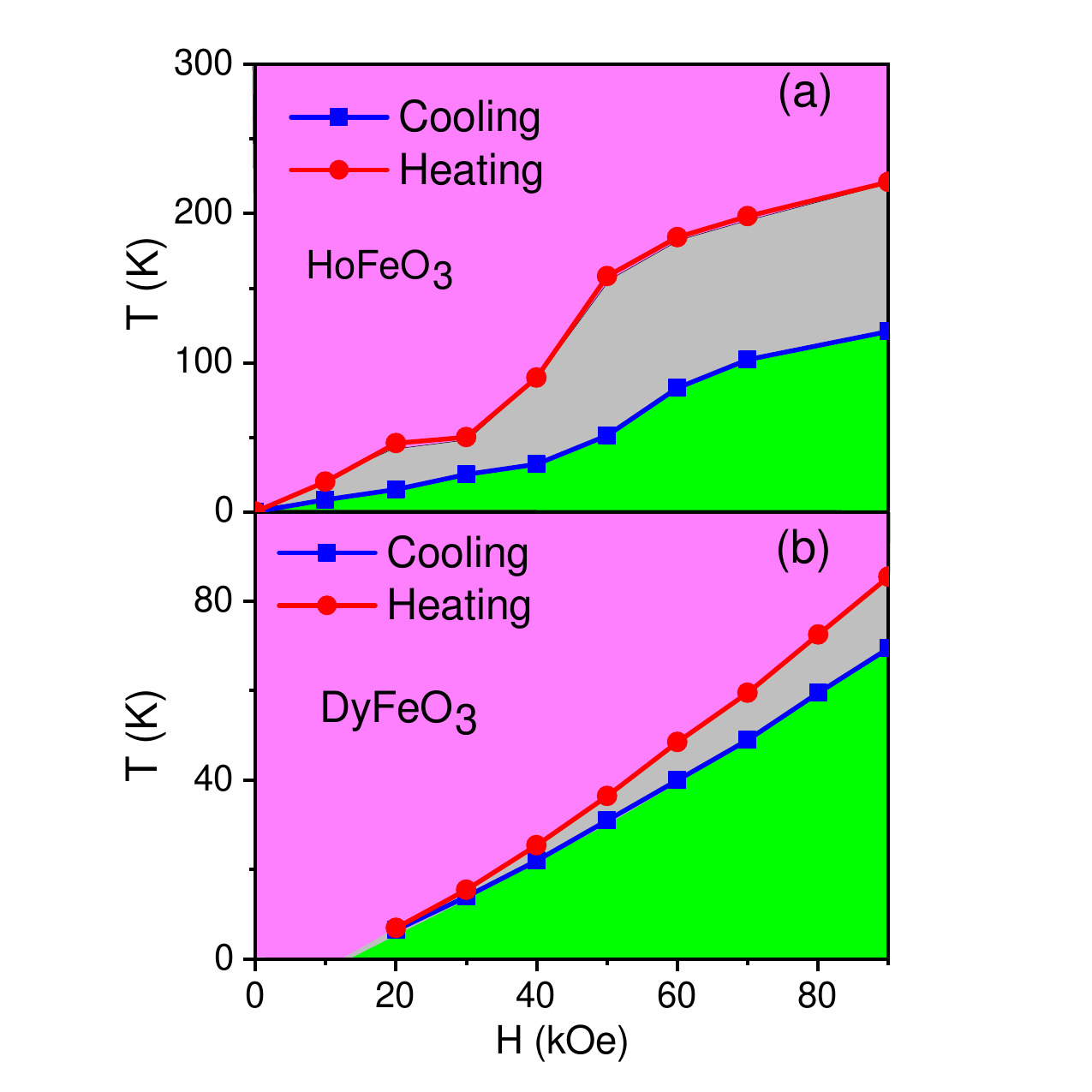}
\caption{\textbf{Magnetic-field--temperature phase diagrams of HoFeO$_3$ and DyFeO$_3$.}
The transition temperatures, $T_S$, extracted from the thermal-expansion measurements are plotted as a function of magnetic field, $H\parallel c$, for (a) HoFeO$_3$ and (b) DyFeO$_3$. The heating and cooling branches are shown separately, revealing pronounced thermal hysteresis associated with the field-induced first-order transition. In DyFeO$_3$, $T_S$ increases approximately linearly with field up to 90 kOe, whereas in HoFeO$_3$ the field dependence is non-linear, with a more rapid increase above approximately 35 kOe. The hysteresis width is generally larger in HoFeO$_3$ than in DyFeO$_3$.}
\label{fig:phase_dia}
\end{figure}

\subsection*{Magnetostructural phase diagram}

The unprecedentedly large values of $c$-axis magnetostriction in HoFeO$_3$ and DyFeO$_3$ indicate that the magnetic order parameter is strongly coupled with the lattice degree of freedom along the $c$ axis. Furthermore, the highly anisotropic nature of magnetostriction suggests that inter-layer Fe-$R$ magnetic interaction plays an important role. Indeed, Raman scattering has demonstrated that the strong spin–phonon coupling leads to a remarkably large shift of phonon frequencies and the appearance of new phonon modes in SmFeO$_3$ \cite{Mads}. The magnetic-field-induced first-order transition has been reported in manganites with crystal structures similar to $R$FeO$_3$ and in several other compounds \cite{Asam,Kuwa,More}. In most of these systems, however, the structural transition already occurs at a finite temperature in zero field and it shifts towards lower or higher temperature upon application of a magnetic field and the value of magnetostriction is very small. Moreover, the reported field-induced strains in several rare-earth transition-metal perovskites are substantially smaller than those observed here. \cite{Rhyne} Figure \ref{fig:phase_dia} shows the structural phase diagram of HoFeO$_3$ and DyFeO$_3$ in the temperature-magnetic field plane with field along the $c$ axis. Both the compounds exhibit strong thermal hysteresis, which increases with increasing field. In DyFeO$_3$,  transition temperature ($T_S$) increases almost linearly with field up to 90 kOe whereas this dependence in HoFeO$_3$ is sensitive to the field range. At low fields, $T_S$ is approximately linear in $H$ but increases rapidly at around 35 kOe.  In general, the width of hysteresis is significantly larger in HoFeO$_3$.

\section{Summary}

We have demonstrated that both HoFeO$_3$ and DyFeO$_3$ exhibit gigantic magnetostriction and a pronounced magnetodielectric response along the $c$ axis, associated with a magnetic-field-induced first-order structural phase transition. The magnetostriction along the other two crystallographic axes is approximately two orders of magnitude smaller and is comparable to that observed in GdFeO$_3$. The strong anisotropy of the response, together with its systematic evolution from Gd to Ho to Dy, points to an important role of rare-earth single-ion anisotropy, magnetocrystalline anisotropy, and magnetic coupling between the rare-earth and Fe sublattices. These results establish rare-earth orthoferrites as a promising platform for exceptionally large, reversible magnetic-field-controlled lattice responses, with potential applications in magnetic actuators, sensors, transducers, adaptive structures and strain-mediated multifunctional devices.

\section{Methods}

\subsection{Crystal growth and structural characterization}

\noindent
Single crystals of $R$FeO$_3$ ($R$ = Ho, Dy, Gd) were grown in a four-mirror image furnace (Crystal Systems Inc.). The image furnace is equipped with four halogen incandescent lamps and hemi-elliptical focusing mirrors. To obtain an oxygen stoichiometry close to 3, the crystals were grown under an oxygen atmosphere at a typical growth rate of 4 mm h$^{-1}$. The phase purity and crystal structure of $R$FeO$_3$ were examined by high-resolution powder x-ray diffraction using Cu K$\alpha$ radiation ($\lambda$ = 1.5406~\AA) on a Rigaku TTRAX II diffractometer at room temperature. All diffraction peaks could be indexed using the distorted orthorhombic structure with $Pbnm$ crystallographic symmetry. The lattice parameters obtained from Rietveld refinement are $a$ = 5.3460~\AA, $b$ = 5.5879~\AA, and $c$ = 7.6680~\AA\ for HoFeO$_3$, and $a$ = 5.3214~\AA, $b$ = 5.5912~\AA, and $c$ = 7.6256~\AA\ for DyFeO$_3$. The obtained values are consistent with the orthorhombic $Pbnm$ structure reported previously.\cite{Chatterji, Biswas} The crystallographic orientations and crystalline quality of the samples were further examined by Laue diffraction. Sharp and well-defined Laue diffraction spots were observed, confirming the high crystalline quality of the single crystals.

For the thermal-expansion and magnetostriction measurements, the crystals were cut into approximately cubic specimens with typical dimensions of $\sim1\times1\times1$ mm$^3$. The crystallographic direction along which the length change was measured was identified by Laue diffraction before mounting the sample in the dilatometer.

\subsection{Magnetic measurements}

Magnetic measurements were performed using a Physical Property Measurement System (PPMS) and a superconducting quantum interference device vibrating-sample magnetometer (SQUID-VSM; Quantum Design). Temperature-dependent magnetization, $M(T)$, was measured at fixed applied magnetic fields using zero-field-cooled (ZFC), field-cooled cooling (FCC), and field-cooled warming (FCW) protocols.

For the ZFC protocol, the sample was cooled from room temperature to the lowest measurement temperature of 2 K in zero magnetic field. The desired magnetic field was then applied, and the magnetization was recorded during warming. For the FCC protocol, the magnetic field was applied at room temperature and the magnetization was recorded while cooling the sample to 2 K. For the FCW protocol, the sample was first cooled to 2 K in the applied magnetic field and the magnetization was subsequently recorded during warming under the same field.

A weak thermal hysteresis near the spin-reorientation temperature, $T_{SR}$, was observed for HoFeO$_3$ and DyFeO$_3$. No corresponding bifurcation was observed for GdFeO$_3$ over the investigated temperature range.

Isothermal magnetization, $M(H)$, was measured at selected fixed temperatures. After completion of each hysteresis-loop measurement, the applied magnetic field was returned to zero, the sample was warmed to room temperature, and subsequently cooled again to the desired measurement temperature before the next $M(H)$ measurement. Five-quadrant magnetic hysteresis loops were recorded for all samples.

\subsection{Magnetostriction and thermal-expansion measurements}

\begin{figure}[ht!]
\centering
\includegraphics[width=\columnwidth]{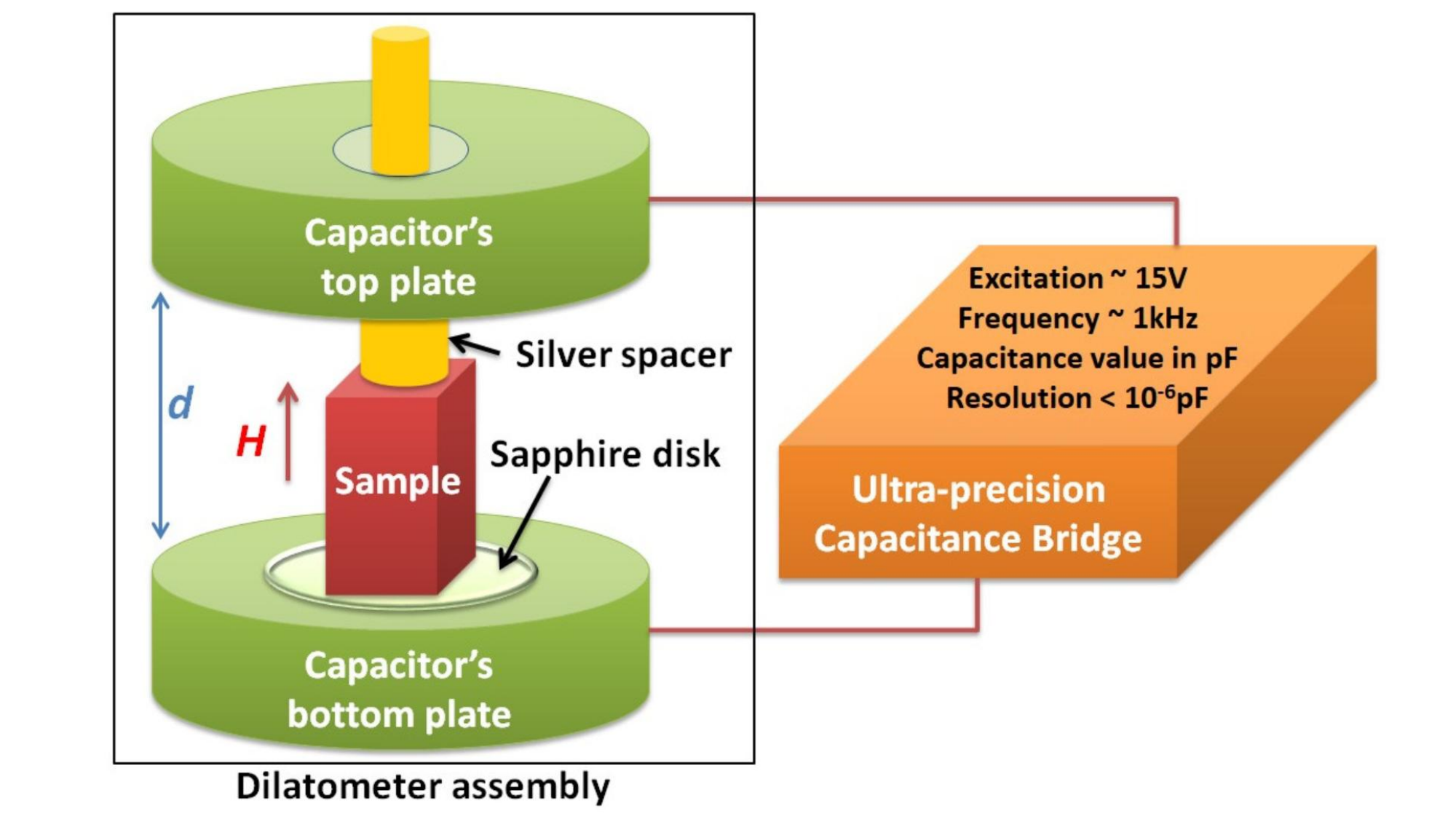}
\caption{Schematic of the capacitive dilatometer and capacitance-bridge configuration used for measurements of temperature- and magnetic-field-induced length changes.}
\label{fig:schematic}
\end{figure}

Temperature- and magnetic-field-dependent length changes were measured using a miniature circular-plate capacitive dilatometer from Vienna Technical University connected to an Andeen-Hagerling AH2700A ultra-precision capacitance and loss bridge operating from 50 Hz to 20 kHz. The sample was mounted between the dilatometer plates, as illustrated in Figure~\ref{fig:schematic}, with a silver spacer placed on the upper surface of the sample to adjust the distance between the plates. The magnetic field, $H$, was applied parallel to the crystallographic axis along which the length change was measured.

An ac excitation of $\sim$15 V at a frequency of $\sim$1 kHz was applied to the capacitor plates. Changes in the capacitance of the dilatometer were converted into changes in the sample length using the calibrated geometrical response of the tilted-plate dilatometer.

For the temperature-dependent measurements, the fractional length change was defined as $\Delta L(T,H)/L_0=[L(T,H)-L_0]/L_0$, where $L_0=L(300~\mathrm{K},0)$ is the sample length at 300 K and zero magnetic field, and $L(T,H)$ is the sample length at temperature $T$ and magnetic field $H$. The percentage length change plotted as thermal expansion $(\Delta l/l)$ was obtained by multiplying $\Delta L/L_0$ by 100. Temperature-dependent length-change measurements were performed between 3 and 300 K at fixed magnetic fields during both cooling and heating. Pronounced thermal hysteresis was observed over the transition region. For field-dependent measurements at a fixed temperature, the longitudinal magnetostriction was defined as $s(T,H)=[L(T,H)-L(T,0)]/L(T,0)$, where $L(T,0)$ is the zero-field sample length at the same measurement temperature. Magnetostriction is reported as $s(H,T)\times100$ in percent.

Before each field-dependent magnetostriction measurement, the magnetic field was returned to zero and the measurement sequence was reset using the same experimental protocol as that used for the $M(H)$ measurements described above. Butterfly-like hysteresis loops were observed at low temperatures, with the hysteresis progressively decreasing as the temperature approached room temperature.

The magnetostriction measurements were repeated at different times over an extended period and showed quantitatively and qualitatively reproducible behaviour. After each measurement cycle, the samples recovered their original dimensions upon returning to room temperature and zero magnetic field. No permanent deformation or mechanical damage of the samples was observed after repeated measurement cycles, despite the exceptionally large field-induced length changes.

The estimated uncertainty in the reported large magnetostriction values is approximately $\pm 2$--$3$\%, dominated by the capacitance-to-displacement calibration, sample-dimension uncertainty, and mechanical reproducibility of the dilatometer.

\subsection{Dielectric and magnetodielectric measurements}

Dielectric measurements were carried out using the same AH2700A ultra-precision capacitance bridge. The dielectric constant was obtained from the measured capacitance as a function of temperature and magnetic field, with the electric probing field applied along the selected crystallographic direction.

\section*{Data availability}
The data that support the findings of this study are available from the corresponding
author upon reasonable request.

\begin{acknowledgments}
The authors thank Arun Paul for assistance with crystal growth and measurements.
\end{acknowledgments}

\section*{Author contributions}
P.M. conceived, designed and supervised the whole project. M.D. synthesized the samples and performed the powder x-ray diffraction and magnetic measurements. M.D. and A.G. performed magnetostriction and magnetodielectric measurements. M.N. and A.T. performed Laue diffraction of single crystals. M.D., A.G. and D.B. analyzed the data. D.B., P.M. and M.D. wrote the initial manuscript. P.M. and A.G. revised and refined the manuscript and prepared the final version for submission. All authors discussed the results and contributed to the final manuscript.

\section*{Competing interests}
The authors declare no conflict of interest.

\section*{Additional information}

\textbf{Supplementary information.}
Supplementary Information is available for this paper and includes x-ray and Laue diffraction characterization of the single crystals; additional representative temperature- and field-dependent magnetization measurements; thermal-expansion and magnetostriction measurements along the $a$ and $b$ crystallographic axes; comparisons of the characteristic fields obtained from magnetization, magnetostriction and dielectric measurements; and magnetostriction measurements of GdFeO$_3$.

\textbf{Correspondence and requests for materials.}
Correspondence and requests for materials should be addressed to the corresponding author: Prabhat Mandal (prabhat.mandal@saha.ac.in).

\end{document}